\documentclass[showpacs]{revtex4}
\usepackage{graphicx}

\begin{document}
\title{Energy of gravitational radiation in plane-symmetric space-times}
\author{Sean A. Hayward}
\affiliation{Center for Astrophysics, Shanghai Normal University, 100 Guilin
Road, Shanghai 200234, China}
\date{19th May 2008}

\begin{abstract}
Gravitational radiation in plane-symmetric space-times can be encoded in a 
complex potential, satisfying a non-linear wave equation. An effective energy 
tensor for the radiation is given, taking a scalar-field form in terms of the 
potential, entering the field equations in the same way as the matter energy 
tensor. It reduces to the Isaacson energy tensor in the linearized, 
high-frequency approximation. An energy conservation equation is derived for a 
quasi-local energy, essentially the Hawking energy. A transverse pressure 
exerted by interacting low-frequency gravitational radiation is predicted. 
\end{abstract}
 \maketitle

\section{Introduction}
Gravitational radiation, as predicted by Einstein gravity, is indirectly 
observed in such examples as the Hulse-Taylor pulsar, and widely expected to be 
directly observed in the coming years, offering a new window to understand 
various astrophysical processes, such as binary inspiral and merger of black 
holes or neutron stars. However, the textbook theory of gravitational radiation 
mostly concerns weak radiation, either in the linearized approximation or at 
infinity in an asymptotically flat space-time \cite{MTW}. Comparatively little 
is known about strong-field radiation. One exception is plane gravitational 
radiation, where exact solutions describe radiation propagating in one 
direction. The simplest scenario to study interaction effects is the head-on 
collision of two such beams, as pioneered by Szekeres \cite{Sz1,Sz2} and 
reviewed by Griffiths \cite{Gri}. More generally, one may study plane symmetric 
space-times, which in vacuum generally consist of gravitational radiation 
propagating in opposite directions and interacting \cite{apw}. 

Much is known about such space-times, including that the interaction is 
non-linear, that the key dynamical equations can be cast as a complex Ernst 
equation \cite{Ern}, and that the cross-focusing of the radiation produces a 
caustic which is generically a curvature singularity, though there are 
non-generic exceptions \cite{CH,cwbh,snc}. This article introduces an effective 
energy tensor $\Theta$ for the gravitational radiation, taking a scalar-field 
form in terms of a complex potential $\Phi$. Then $\Theta$ enters the field 
equations in the same way as the matter energy tensor, in particular entering 
an energy conservation law. The Ernst equation is manifestly a wave equation 
for $\Phi$, generally with a non-linear source, which vanishes for collinear 
polarization. 

The method involves a conserved time vector $k^a$, a conserved energy-momentum 
density $j^a$, a corresponding energy $E$ and a first law for $E$ involving 
energy-supply and work terms. Surface gravity $\kappa$ is also defined and 
takes a quasi-Newtonian form. This is intended to complete the same programme 
of identifying physical quantities and equations which has previously been 
performed in spherical symmetry \cite{sph,1st}, cylindrical symmetry \cite{cyl} 
and a quasi-spherical approximation \cite{qs,SH,gwbh,gwe}. These references 
will be assumed for comparison throughout the text without repeated citation, 
though the treatment here is self-contained. 

\section{Metric variables and field equations}
Cartesian coordinates $(z,y)$ on the planes of symmetry will be used, to allow 
easy comparisons with standard coordinates $(z,\varphi)$ in cylindrical 
symmetry and $(\vartheta,\varphi)$ in spherical symmetry and the 
quasi-spherical approximation. It is convenient to use null coordinates $x^\pm$ 
in the normal space, as they are adapted to gravitational radiation. Then the 
metric can be written locally as 
\begin{equation}
ds^2=-2e^{2\gamma}dx^+dx^-+A\left(e^{2\phi}\sec2\chi dy^2+2\tan2\chi
dydz+e^{-2\phi}\sec2\chi dz^2\right) \label{met}
\end{equation}
where $(A,\phi,\chi,\gamma)$ are functions of $(x^+,x^-)$. Here $A$ is the 
specific area, meaning that it is the area of a square coordinate patch 
$(0,1)\times(0,1)$ in the $(y,z)$ plane. It is invariant up to constant linear 
transformations of $y$ and $z$, under which it scales by a constant factor. The 
remaining freedom in $(y,z)$ is by rotations, under which $A$ is invariant. The 
functions $(\phi,\chi)$ encode the gravitational radiation, as will be seen 
below. They are invariant up to the above-mentioned transformations of $(y,z)$, 
which will be treated as fixed henceforth. The remaining function $\gamma$ is 
invariant up to functional rescalings $x^\pm\mapsto\tilde x^\pm(x^\pm)$, under 
which it transforms by additive functions of $x^+$ and $x^-$. The variables 
have been chosen so that the induced metric on the planes of symmetry takes a 
similar form to that used in the quasi-spherical approximation, with $(dz,dy)$ 
replaced by $(d\vartheta,\sin\vartheta d\varphi)$, and takes a similar form to 
that used in cylindrical symmetry. The Szekeres variables $(P,M,Q,W)$ are 
related by 
\begin{equation}
P=-\log A,\quad M=-2\gamma,\quad Q=-2\phi,\quad\sinh W=\tan2\chi  \label{sz}
\end{equation}
or $\cosh W=\sec2\chi$.

The six independent components of the Einstein equation may be found directly,
or by comparison with the Szekeres form, as
\begin{eqnarray}
&&2A\partial_\pm\partial_\pm A-(\partial_\pm A)^2 -4A\partial_\pm 
A\partial_\pm\gamma+4A^2\sec^22\chi\left((\partial_\pm\phi)^2+(\partial_\pm\chi)^2\right) 
=-16\pi A^2T_{\pm\pm}\label{ei}\\
&&\partial_+\partial_-A=8\pi AT_{+-}\label{ea}\\
&&2A\partial_+\partial_-\phi+\partial_+A\partial_-\phi+\partial_-A\partial_+\phi
+4A\tan2\chi(\partial_+\phi\partial_-\chi+\partial_-\phi\partial_+\chi)
=4\pi Ae^{2\gamma}\cos^22\chi(T^y_y-T^z_z)\label{ep}\\
&&2A\partial_+\partial_-\chi+\partial_+A\partial_-\chi+\partial_-A\partial_+\chi
+4A\tan2\chi(\partial_+\chi\partial_-\chi-\partial_+\phi\partial_-\phi)
=4\pi Ae^{2\gamma}\cos^22\chi(e^{2\phi}T^y_z+e^{-2\phi}T^z_y)\label{ec}\\
&&4A^2\partial_+\partial_-\gamma-\partial_+A\partial_-A
+4A^2\sec^22\chi(\partial_+\phi\partial_-\phi+\partial_+\chi\partial_-\chi)
=-8\pi A^2\left(2T_{+-}+e^{2\gamma}(T^y_y+T^z_z)\right)\label{eg}
\end{eqnarray}
where $\partial_\pm=\partial/\partial x^\pm$, $T$ denotes the energy tensor of
the matter with $T_{\pm\pm}=T(\partial_\pm,\partial_\pm)$,
$T_{+-}=T(\partial_+,\partial_-)$, and the units are such that Newton's
gravitational constant is unity. The equations (\ref{ei}) can be regarded as
constraint equations on initial null hypersurfaces $\Sigma_\pm$ of constant
$x^\mp$, as they are preserved in the $\partial_\mp$ directions due to the
Bianchi identities or energy-momentum conservation. The other equations
(\ref{ea})--(\ref{eg}) are then the evolution equations.

In vacuum, $T=0$, it is well known that these equations describe the 
propagation and interaction of gravitational radiation in the opposite 
$\partial_\pm$ directions, and that the radiation may be encoded in 
$(\phi,\chi)$. The solution to (\ref{ea}) is trivial and can be used to fix the 
rescaling freedom in $x^\pm$. One may give initial data for $(\phi,\chi)$ on 
$\Sigma_\pm$, corresponding to initial radiation profiles, with (\ref{ei}) 
determining $\gamma$ on $\Sigma_\pm$. Then the main task is to solve 
(\ref{ep})--(\ref{ec}) simultaneously for $(\phi,\chi)$, after which the full 
solution follows from (\ref{eg}) by quadrature for $\gamma$. The main equations 
(\ref{ep})--(\ref{ec}) can be written as a complex Ernst equation, 
corresponding physically to a non-linear wave equation, as will be verified 
below. 

\section{Effective energy tensor for gravitational radiation} 
The next aim is to find an effective energy tensor $\Theta$ for the 
gravitational radiation, analogous to those found in cylindrical symmetry and 
the quasi-spherical approximation, and consistent with the Isaacson effective 
energy tensor in the high-frequency linearized approximation \cite{MTW}. In all 
cases, the components of the energy tensor are quadratic in first derivatives 
of the metric, in this case the $\partial_\pm$ derivatives of $(\phi,\chi)$, 
and such terms can be seen in the last term in parentheses on the left-hand 
side of each of (\ref{ei}), (\ref{ep})--(\ref{eg}). The idea is to identify 
these terms as components of the desired $\Theta$, corresponding to the 
components of $T$ on the right-hand sides. The result is that one may introduce 
a complex potential 
\begin{equation}
\Phi=\phi+i\chi \label{Ph}
\end{equation}
and define the effective energy tensor as
\begin{equation}
\Theta_{ab}=\frac{2\nabla_{(a}\Phi\nabla_{b)}\bar\Phi
-g_{ab}g^{cd}\nabla_c\Phi\nabla_d\bar\Phi}
{8\pi\cosh^2(\Phi-\bar\Phi)}.\label{Th}
\end{equation}
where $g$ is the space-time metric and $\nabla$ its covariant derivative 
operator. It is manifestly a tensor, taking a scalar-field form in terms of 
$\Phi$, with the same form, including the same denominator, as in the 
quasi-spherical approximation. Apart from this denominator, it is the energy 
tensor of a massless complex scalar field $\Phi$. Explicitly in terms of 
$(\phi,\chi)$, 
\begin{equation}
\Theta_{ab}=\frac
{2\nabla_a\phi\nabla_b\phi+2\nabla_a\chi\nabla_b\chi
-g_{ab}g^{cd}(\nabla_c\phi\nabla_d\phi+\nabla_c\chi\nabla_d\chi)}
{8\pi\cos^22\chi}.
\end{equation}
If $\chi=0$, it reduces to the energy tensor of a massless scalar field $\phi$, 
as in cylindrical symmetry, where the corresponding $\phi$ reduces to the 
Newtonian gravitational potential in the Newtonian limit. Here there are 
generally two polarizations of the radiation, as is familiar from the 
linearized approximation. Inspection of the metric (\ref{met}) for small $\Phi$ 
identifies $\phi$ as encoding the ``plus'' polarization and $\chi$ as encoding 
the ``cross'' polarization. These properties justify the numerical factors 
chosen in the definitions of $(\phi,\chi)$ and partly motivated the chosen 
symbols. 

The non-trivial components of $\Theta$ follow explicitly as
\begin{eqnarray}
&&4\pi\Theta_{\pm\pm}=\sec^22\chi
\left((\partial_\pm\phi)^2+(\partial_\pm\chi)^2\right)\label{Th1}\\
&&\Theta_{+-}=0\label{Th0}\\
&&4\pi\bot\Theta=e^{-2\gamma}\sec^22\chi
(\partial_+\phi\partial_-\phi+\partial_+\chi\partial_-\chi)\bot g\label{Th2}
\end{eqnarray}
where $\bot$ denotes projection onto the planes of symmetry and the transverse
metric is given in $(y,z)$ coordinates by
\begin{equation}
\bot g=A\pmatrix{e^{2\phi}\sec2\chi&\tan2\chi\cr\tan2\chi&e^{-2\phi}\sec2\chi}.
\label{h}
\end{equation}
It is then straightforward to verify that adding $\Theta$ to $T$ on the 
right-hand sides of the Einstein equations (\ref{ei})--(\ref{eg}) cancels the 
quadratic terms in $(\phi,\chi)$ on the left-hand sides. In abstract terms, the 
Einstein equation $G=8\pi T$ may be rewritten as $C=8\pi(T+\Theta)$ in terms of 
a truncated Einstein tensor $C$, whose components have a simpler form to those 
of the Einstein tensor $G$. 

The physical interpretation of $\Theta_{\pm\pm}/2$ is the energy density of 
gravitational radiation propagating in the $\partial_\mp$ direction. Apart from 
the non-linear modification due to the $\sec^22\chi$ factor in (\ref{Th1}), it 
is the energy density of a complex scalar field $\Phi$. The numerical factor 
also corresponds to the energy density of electromagnetic radiation in Gaussian 
units, with $\phi$ corresponding to the electric potential and $\chi$ 
vanishing. The vanishing of $\Theta_{+-}$ (\ref{Th0}) is familiar from 
cylindrical symmetry and the quasi-spherical approximation, and indicates that 
the gravitational radiation is workless. Note that this is generally not so for 
a similar effective energy tensor found in the context of black holes 
\cite{bhd2,bhd3} and uniformly expanding flows \cite{gr,BHMS}. The 
non-negativity of $\Theta_{\pm\pm}$ indicates that, as an energy tensor, 
$\Theta$ satisfies the dominant energy condition, meaning physically that 
gravitational radiation carries positive energy. The other non-zero terms 
(\ref{Th2}) indicate that interacting gravitational radiation generally exerts 
transverse pressure and shear, proportional to the transverse metric. These 
terms vanish for radiation propagating in one direction only, where $\Phi$ is a 
function of $x^+$ (or $x^-$) only. They are commonly known as plane waves, but 
since this would appear to imply periodicity in some sense, this article uses 
the more general terminology of radiation. 

\section{Conservation of energy}
To see how $\Theta$ further qualifies as an effective energy tensor, one may 
proceed by analogy with spherical symmetry, cylindrical symmetry and the 
quasi-spherical approximation. Here the definitions and equations will be 
stated first in a manifestly invariant way, then verified in coordinates. First 
introduce the specific area radius 
\begin{equation}
r=\sqrt{A/4\pi}.\label{r}
\end{equation}
This is defined in order to compare with spherically symmetric space-times or 
the quasi-spherical approximation, so that one may easily treat astrophysical 
gravitational radiation as observed on or near Earth, since distant sources can 
be treated as points, producing roughly spherical wavefronts which can be 
treated as planes when observed.

The Hodge operator $*$ defines the Hodge dual ${*}\alpha$ of a normal one-form, 
up to sign, by  
\begin{equation}
g^{-1}({*}\alpha,\alpha)=0,
\quad g^{-1}({*}\alpha,{*}\alpha)=-g^{-1}(\alpha,\alpha).\label{dual}
\end{equation}
Then a preferred time vector is defined by
\begin{equation}
k=g^{-1}({*}dr) \label{k}
\end{equation}
where the qualification ``specific'' is omitted here and henceforth. This 
vector is conserved: 
\begin{equation}
\nabla\cdot k=0. \label{nk}
\end{equation}
The corresponding energy-momentum density is
\begin{equation}
j=-g^{-1}((T+\Theta)\cdot k). \label{j}
\end{equation}
Then $j$ is also conserved:
\begin{equation}
\nabla\cdot j=0. \label{nj}
\end{equation}
Here the standard physical interpretation is conservation of energy, and the 
role of $\Theta$ as an effective energy tensor is clear in that it appears 
additively with $T$ in $j$. 

Put another way, both $k$ and $j$ are Noether currents, and the corresponding 
Noether charges are area volume 
\begin{equation}
V=\textstyle{\frac43}\pi r^3\label{V}
\end{equation}
and energy $E$, defining the latter. Specifically: 
\begin{equation}
Ag(k)={*}dV,\qquad Ag(j)={*}dE. \label{noe}
\end{equation}
Integrating for $E$ and requiring it to vanish for flat space-time,
\begin{equation}
E=-\textstyle{\frac12}rg^{-1}(dr,dr) \label{E}
\end{equation}
which has a similar form to the Misner-Sharp energy in spherical symmetry and 
the modified Thorne energy in cylindrical symmetry. In fact, if the planes of 
symmetry are toroidally compacted by periodic identifications in $(y,z)$ at 0 
and 1, so that $A$ is the area, then $E$ coincides with the Hawking energy 
\cite{Haw}. 

Note that $E>0$ for trapped surfaces, $E=0$ for marginal surfaces and $E<0$ for 
untrapped surfaces. In particular, $E$ vanishes for radiation propagating in 
one direction only. Thus it should not be interpreted as the energy of a wave 
in any sense. Taking the example of two colliding beams, where the surfaces in 
the interaction region are trapped if the null energy condition holds, one may 
interpret $E$ as measuring energy due to cross-focusing of radiation. In 
particular, it diverges at the caustic formed by such cross-focusing. 

Introduce the work density
\begin{equation}
w=-\hbox{tr}\,T/2 \label{w}
\end{equation}
and the energy flux
\begin{equation}
\psi=(T+\Theta)\cdot g^{-1}(dr)+wdr \label{psi}
\end{equation}
where the trace is in the normal space. Then conservation of energy (\ref{nj}) 
can be written in the form of a first law: 
\begin{equation}
dE=A\psi+wdV \label{dE}
\end{equation}
which has the same form as in spherical symmetry and the quasi-spherical 
approximation. Here the two terms can be interpreted as energy supply and work 
respectively, as in the first law of thermodynamics. Note again that $\Theta$ 
appears additively with $T$ in $\psi$ and (in a null sense) $w$, playing the 
role of an effective energy tensor. 

The corresponding definition of surface gravity is 
\begin{equation}
\kappa={*}d{*}dr/2 \label{kappa}
\end{equation}
where $d$ is the exterior derivative of the normal space. Then the Einstein 
equations yield 
\begin{equation}
\kappa=\frac{E}{r^2}-4\pi rw \label{sg}
\end{equation}
which again has the same form as that in spherical symmetry and the 
quasi-spherical approximation. Apart from the matter term, this has the form of 
Newtonian gravitational acceleration. 

In dual-null coordinates (\ref{met}), the corresponding expressions are
\begin{eqnarray}
&&{*}\alpha=-\alpha_+dx^++\alpha_-dx^-\quad\hbox{where $\alpha=\alpha_+dx^++\alpha_-dx^-$}\\
&&k=e^{-2\gamma}(\partial_+r\partial_--\partial_-r\partial_+)\\
&&j=e^{-4\gamma}\big[\big((T_{--}+\Theta_{--})\partial_+r-T_{+-}\partial_-r\big)\partial_+
-\big((T_{++}+\Theta_{++})\partial_-r-T_{+-}\partial_+r\big)\partial_-\big]\\
&&E=e^{-2\gamma}r\partial_+r\partial_-r\\
&&w=e^{-2\gamma}T_{+-}\\
&&\psi_\pm=-e^{-2\gamma}(T_{\pm\pm}+\Theta_{\pm\pm})\partial_\mp r\\
&&\kappa=-e^{-2\gamma}\partial_+\partial_-r.
\end{eqnarray}
Writing $4(4\pi)^{3/2}E=e^{-2\gamma}A^{-1/2}\partial_+A\partial_-A$ and using 
the Einstein equations (\ref{ei})--(\ref{ea}), a calculation yields 
\begin{equation}
\partial_\pm E=Ae^{-2\gamma}\big(\partial_\pm rT_{+-}
-\partial_\mp r(T_{\pm\pm}+\Theta_{\pm\pm})\big).
\end{equation}
Comparison with
\begin{eqnarray}
Ag(j)&=&Ae^{-2\gamma}\big[\big((T_{++}+\Theta_{++})\partial_-r-T_{+-}\partial_+r\big)dx^+
-\big((T_{--}+\Theta_{--})\partial_+r-T_{+-}\partial_-r\big)dx^-\big] \nonumber\\
&=&[-\partial_+Edx^++\partial_-Edx^-]={*}dE
\end{eqnarray}
verifies (\ref{noe}). Similarly, the calculation
\begin{equation}
A(\psi_\pm+w\partial_\pm r)
=Ae^{-2\gamma}\big(-\partial_\mp r(T_{\pm\pm}+\Theta_{\pm\pm})+\partial_\pm rT_{+-}\big)
=\partial_\pm E
\end{equation}
verifies (\ref{dE}). The easiest way to verify the conservation equations 
(\ref{nk}), (\ref{nj}) is to use (\ref{noe}) and exterior calculus: 
\begin{eqnarray}
\nabla\cdot k&=&A^{-1}{*}d{*}(Ag(k))=A^{-1}{*}d{*}{*}dV=0\\
\nabla\cdot j&=&A^{-1}{*}d{*}(Ag(j))=A^{-1}{*}d{*}{*}dE=0
\end{eqnarray}
since ${*}{*}=\pm1$ and $dd=0$. Finally, a calculation using the Einstein 
equation (\ref{ea}) verifies (\ref{sg}). 

\section{Gravitational wave equation}
As is well known, the propagation equations (\ref{ep})--(\ref{ec}) for 
$(\phi,\chi)$ can be written as a single complex Ernst equation, usually given 
in terms of an Ernst potential $Z=e^{2\Phi}$ or $E=\tanh\Phi$ \cite{Gri}. The 
corresponding form for $\Phi$ is 
\begin{equation}
\nabla^2\Phi=2\tanh(\Phi-\bar\Phi)g^{-1}(\nabla\Phi,\nabla\Phi)\label{ernst}
\end{equation}
where $\bot T=0$ for simplicity. This has the same form as that in the 
quasi-spherical approximation. The calculation is straightforward: 
\begin{equation}
\nabla^2\Phi=-e^{-2\gamma}\left(2\partial_+\partial_-\Phi
+A^{-1}(\partial_+A\partial_-\Phi+\partial_-A\partial_+\Phi)\right)
\end{equation}
and
\begin{eqnarray}
2\tanh(\Phi-\bar\Phi)g^{-1}(\nabla\Phi,\nabla\Phi)
&=&-4e^{-2\gamma}\tanh2i\chi(\partial_+\phi+i\partial_+\chi)(\partial_-\phi+i\partial_-\chi) \nonumber\\
&=&4e^{-2\gamma}\tan2\chi\left((\partial_+\phi\partial_-\chi+\partial_-\phi\partial_+\chi)
+i(\partial_+\chi\partial_-\chi-\partial_+\phi\partial_-\phi)\right)
\end{eqnarray}
then the result follows by comparing with (\ref{ep})--(\ref{ec}).

Note that (\ref{ernst}) is manifestly a wave equation for $\Phi$, equating 
$\nabla^2\Phi$ to a non-linear term in $\Phi$. This source term is highly 
non-linear, being quadratic in $\nabla\Phi$ and also involving 
$\tanh(\Phi-\bar\Phi)$. In the special case of collinear polarization $\chi=0$, 
the source term vanishes and the equation reduces to the wave equation for 
$\phi$, $\nabla^2\phi=0$. This can be written as an Euler-Poisson-Darboux 
equation, for which general solutions are available. The full Ernst equation 
has been studied by various methods both in plane symmetry and in the original 
context of stationary axisymmetric space-times; see e.g.\ the review of 
Griffiths \cite{Gri} and references therein. 

\section{Linearized gravitational radiation}
To compare with the usual description of linearized gravitational radiation 
\cite{MTW}, it is convenient to switch temporarily to Minkowski coordinates 
$(t,x,y,z)$ defined by $\sqrt2 x^\pm=t\pm x$. Expanding about the Minkowski 
metric $\eta=\hbox{diag}\{-1,1,1,1\}$ by $g=\eta+h$ consists of expanding about 
$(A,\phi,\chi,\gamma)=(1,0,0,0)$, so one can write $A=1+\alpha$ and use  
$(\alpha,\phi,\chi,\gamma)$ as perturbative fields, each assumed ${}\ll1$. 
Linearizing, the metric perturbation $h$ is given by 
\begin{equation}
-2\gamma(dt^2-dx^2)+(\alpha+2\phi)dy^2+2(\alpha+2\chi)dydz+(\alpha-2\phi)dz^2.
\end{equation}
Then the trace of $h$ is $2\alpha+4\gamma$ and the trace-reversed metric 
perturbation $\bar h$ is given by 
\begin{equation}
\alpha(dt^2-dx^2)+(2\phi-2\gamma)dy^2+2(\alpha+2\chi)dydz+(-2\phi-2\gamma)dz^2.
\end{equation}
Applying the transverse traceless gauge conditions, $\partial^a\bar h_{ab}=0$ 
yields constant $\alpha$, $\bar h_{0b}=0$ yields $\alpha=0$ and $\bar h^a_a=0$ 
yields $\gamma=0$. Then $h=\bar h$ is indeed transverse: in $(y,z)$ 
coordinates, 
\begin{equation}
h=\pmatrix{2\phi&2\chi\cr2\chi&-2\phi}.
\end{equation}
This verifies the appropriateness of the transverse traceless gauge conditions 
in plane symmetry. Noting that the space-time strain is $h/2$, this also 
confirms that $\phi$ and $\chi$ encode the ``plus'' and ``cross'' polarizations 
respectively. 

In the high-frequency approximation, the Isaacson effective energy tensor 
$\bar\Theta$ for gravitational waves is defined by 
\begin{equation}
32\pi\bar\Theta_{ab}=
\langle\partial_a h_{cd}\partial_bh^{cd}\rangle
\end{equation}
where the angle brackets denote averaging over several wavelengths \cite{MTW}. 
Returning to dual-null coordinates, the explicit expressions are 
\begin{eqnarray}
&&4\pi\bar\Theta_{\pm\pm}=\langle(\partial_\pm\phi)^2+(\partial_\pm\chi)^2\rangle\\
&&4\pi\bar\Theta_{+-}=\langle\partial_+\phi\partial_-\phi+\partial_+\chi\partial_-\chi\rangle\\
&&\bot\bar\Theta=0.
\end{eqnarray}
Comparing with (\ref{Th1}--\ref{Th2}), one sees that the radiative components 
$\bar\Theta_{\pm\pm}$ agree with $\Theta_{\pm\pm}$, but the other components 
apparently do not. However, this is due to the averaging, as follows.

First note that the gravitational wave equation (\ref{ernst}) linearizes to the 
flat-space form 
\begin{equation}
\partial_+\partial_-\Phi=0
\end{equation}
with general solution
\begin{equation}
\Phi=\Phi_+(x^+)+\Phi_-(x^-)
\end{equation}
as expected. Considering linear superpositions of Fourier modes in the 
high-frequency approximation, it suffices to consider solutions of the form 
\begin{equation}
\Phi_\pm=\phi_\pm\sin\sqrt2\omega_\pm x^\pm
+i\chi_\pm\sin\sqrt2\nu_\pm x^\pm
\end{equation}
for constant amplitudes $(\phi_\pm,\chi_\pm)$ and angular frequencies 
$(\omega_\pm,\nu_\pm)$. Then 
\begin{eqnarray}
&&\partial_\pm\phi=\sqrt2\phi_\pm\omega_\pm\cos\sqrt2\omega_\pm x^\pm\\
&&\partial_\pm\chi=\sqrt2\chi_\pm\nu_\pm\cos\sqrt2\nu_\pm x^\pm
\end{eqnarray}
and 
\begin{eqnarray}
&&4\pi\Theta_{\pm\pm}=2\phi_\pm^2\omega_\pm^2\cos^2\sqrt2\omega_\pm x^\pm
+2\chi_\pm^2\nu_\pm^2\cos^2\sqrt2\nu_\pm x^\pm\\
&&4\pi\bot\Theta
=2\left(\phi_+\phi_-\omega_+\omega_-\cos\sqrt2\omega_+x^+\cos\sqrt2\omega_-x^-
+\chi_+\chi_-\nu_+\nu_-\cos\sqrt2\nu_+x^+\cos\sqrt2\nu_-x^-\right)\delta
\end{eqnarray}
where $\delta=\hbox{diag}\{1,1\}$. Since $\langle\cos^2\rangle=\frac12$ but 
$\langle\cos\rangle=0$, $\langle\bot\Theta\rangle=0$ and similarly 
$\bar\Theta_{+-}=0$. Then 
\begin{eqnarray}
&&4\pi\langle\Theta_{\pm\pm}\rangle=4\pi\bar\Theta_{\pm\pm}
=\phi_\pm^2\omega_\pm^2+\chi_\pm^2\nu_\pm^2\\
&&\langle\Theta_{+-}\rangle=\bar\Theta_{+-}=0\\
&&\langle\bot\Theta\rangle=\bot\bar\Theta=0
\end{eqnarray}
or
\begin{equation}
\langle\Theta\rangle=\bar\Theta
\end{equation}
as expected. Note that the energy densities $\bar\Theta_{\pm\pm}/2$ have the 
expected form of squares of amplitudes times angular frequencies, with the same 
numerical factor $1/8\pi$ as for electromagnetic radiation in Gaussian units. 

On the other hand, for low-frequency waves, transverse pressure is generally 
present in $\bot\Theta$ even in the linearized approximation, for which 
$\Theta$ reduces to the energy tensor of a massless complex scalar field in 
flat space-time: 
\begin{equation}
8\pi\Theta_{ab}=2\partial_{(a}\Phi\partial_{b)}\bar\Phi
-\eta_{ab}\eta^{cd}\partial_c\Phi\partial_d\bar\Phi.
\end{equation}
The non-zero components (\ref{Th1})--(\ref{Th2}) reduce to 
\begin{eqnarray}
&&4\pi\Theta_{\pm\pm}=(\partial_\pm\phi)^2+(\partial_\pm\chi)^2\\
&&4\pi\bot\Theta=
(\partial_+\phi\partial_-\phi+\partial_+\chi\partial_-\chi)\delta
\end{eqnarray}
and in particular the transverse shear vanishes, but transverse pressure 
generally remains. Recall that this is an effect for interacting radiation, 
vanishing for radiation propagating in one direction only. However, if two 
beams with similar amplitude and frequency are passing through one another, the 
transverse pressure is generally of the same order as the energy densities 
$\Theta_{\pm\pm}/2$. Although this has been derived here only for 
plane-symmetric radiation propagating in opposite directions, one may expect it 
to generalize to gravitational radiation from any two sources in different 
directions. 

\medskip 

Research supported by the National Natural Science Foundation of China under 
grants 10375081, 10473007 and 10771140, by Shanghai Municipal Education 
Commission under grant 06DZ111, and by Shanghai Normal University under grant 
PL609.


\begin{thebibliography}{99}
\bibitem{MTW}C W Misner, K S Thorne \& J A Wheeler,
 Gravitation (Freeman 1973).
\bibitem{Sz1}P Szekeres, Nature {\bf 228}, 1183. 
\bibitem{Sz2}P Szekeres, {J. Math. Phys.} {\bf 13}, 286 (1972). 
\bibitem{Gri}J B Griffiths, Colliding Waves in General Relativity (Oxford 
    University Press 1991). 
\bibitem{apw}S A Hayward, {Class. Quantum Grav.} {\bf 7}, 1117 (1990).
\bibitem{Ern}F J Ernst, {Phys. Rev.} {\bf 167}, 1175 (1968).
\bibitem{CH}C J S Clarke \& S A Hayward, {Class. Quantum Grav.} {\bf 6}, 615 
    (1989). 
\bibitem{cwbh}S A Hayward, {Class. Quantum Grav.} {\bf 6}, 1021 (1989).
\bibitem{snc}S A Hayward, {Class. Quantum Grav.} {\bf 6}, L179 (1989).
\bibitem{sph}S A Hayward, {Phys. Rev.} {\bf D53}, 1938 (1996).
\bibitem{1st}S A Hayward, {Class. Quantum Grav.} {\bf 15}, 3147 (1998).
\bibitem{cyl}S A Hayward, {Class. Quantum Grav.} {\bf 17}, 1749 (2000). 
    Corrigendum ibid 4159. 
\bibitem{qs}S A Hayward, {Phys. Rev.} {\bf D61}, 101503 (2000).
\bibitem{SH}H Shinkai \& S A Hayward, {Phys. Rev.} {\bf D64}, 044002 (2001). 
\bibitem{gwbh}S A Hayward, {Class. Quantum Grav.} {\bf 18}, 5561 (2001).
\bibitem{gwe}S A Hayward, {Phys. Lett.} {\bf A294}, 179 (2002).
\bibitem{bhd2}S A Hayward, {Phys. Rev. Lett.} {\bf 93}, 251101 (2004).
\bibitem{bhd3}S A Hayward, {Phys. Rev.} {\bf D70}, 104027 (2004).
\bibitem{gr}S A Hayward, {Class. Quantum Grav.} {\bf 23}, L15 (2006).
\bibitem{BHMS}H Bray, S A Hayward, M Mars \& W Simon, {Comm. Math. Phys. 
    Lett.} {\bf 272}, 119 (2007). 
\bibitem{Haw}S W Hawking, {J. Math. Phys.} {\bf 9}, 598 (1968).
\end{thebibliography}
\end{document}